# Multi-Microgrid Collaborative Optimization Scheduling Using an Improved Multi-Agent Soft Actor-Critic Algorithm

Jiankai Gao [1], Yang Li [1,*], Bin Wang [2] and Haibo Wu [1]

[1] School of Electrical Engineering, Northeast Electric Power University, Jilin 132012, China
[2] State Grid Jining Power Supply Company, Jining 272000, China
* Correspondence: liyang@neepu.edu.cn.

**Abstract:** The implementation of a multi-microgrid (MMG) system with multiple renewable energy sources enables the facilitation of electricity trading. To tackle the energy management problem of a MMG system, which consists of multiple renewable energy microgrids belonging to different operating entities, this paper proposes a MMG collaborative optimization scheduling model based on a multi-agent centralized training distributed execution framework. To enhance the generalization ability of dealing with various uncertainties, we also propose an improved multi-agent soft actor-critic (MASAC) algorithm, which facilitates energy transactions between multi-agents in MMG, and employs automated machine learning (AutoML) to optimize the MASAC hyperparameters to further improve the generalization of deep reinforcement learning (DRL). The test results demonstrate that the proposed method successfully achieves power complementarity between different entities, and reduces the MMG system operating cost. Additionally, the proposal significantly outperforms other state-of-the-art reinforcement learning algorithms with better economy and higher calculation efficiency.

## 1. Introduction

To achieve sustainable social development, it is imperative to embrace clean, low-carbon, and sustainable energy sources [1, 2]. However, due to the inherent uncertainty of renewable energy, the integration of multiple renewable energy sources in the form of microgrid (MG) has played a significant role in promoting the consumption of renewable energy [3, 4, 5]. As technology advances, connecting multiple microgrids (MGs) within the same power distribution area can unlock the potential of various flexible resources, enabling the complementary utilization of multi-microgrid (MMG) energy [6]. In addition, this approach further promotes the consumption of various renewable energy sources, which has emerged as a new trend in development [7, 8]. However, the energy interaction between multiple MGs involves complex transaction relationships, leading to significant challenges in system regulation. In this case, it is of great significance to investigate the collaborative optimal dispatch of MMG with electric energy interaction to fully exploit the potential of renewable energy sources and ensure efficient system regulation.

Existing research has made significant progress in addressing the complexity of managing MMG energy. Ref. [9] proposes optimal scheduling of MMG based on federated learning and reinforcement learning. Ref. [10] constructs a MMG system in cold and hot power areas taking into account electric energy interaction. Although the above works have addressed the power interaction in a multi-microgrid system, the benefits of each MG are not considered enough. Regarding the aforementioned issue, some works have addressed the complexity of energy transactions between different entities in MMG systems. Ref. [11] considers an incompletely rational peer-to-peer MG energy transaction. Ref. [12] uses particle swarm optimization (PSO) optimization algorithm for peer-to-peer MMG economic dispatch. Ref. [13] proposes a MMG distributed power management in the shipping area based on an alternating direction method of multipliers (ADMM) algorithm. Ref. [14] leverages Monte Carlo simulations for energy trading of MMG. Nevertheless, the high uncertainty, wide variability, and multi-energy coupling information of MMG systems present significant challenges in modeling the energy transactions between different entities. Currently, there are two main categories of methods for solving the MMG energy management scheduling model: model-driven approaches and data-driven approaches. Studies focusing on model-driven methods have also been conducted in this area. Ref. [15] proposes an improved genetic algorithm for MMG economic dispatch. Ref. [16] utilizes a PSO algorithm for the optimal scheduling of MG containing electric vehicles. Ref. [17] uses a distributed control method for the energy scheduling of MMG. Ref. [18] leverages the ADMM algorithm for the day-ahead scheduling of MMG based on the cooperative game model.

Despite the progress made in this area, research on MMG energy management still faces several challenges due to the complexity of energy transactions between different entities and the uncertainty associated with renewable energy output. These challenges include: (1) the solution methods used heavily rely on the accuracy of the MMG model and lack robustness to the uncertainties associated with multiple available energy sources. Additionally, these methods may consume a significant amount of resources; (2) moreover, the existing solution methods focus primarily on short-term benefits, neglecting the potential long-term benefits. Consequently, finding effective ways to address these challenges has become a key issue in MMG energy management.

To address the above challenges, we propose a data-driven approach that leverages a deep reinforcement learning (DRL) algorithm to coordinate the energy management of MMG. To be specific, deep neural networks avoid the dependence on precise mathematical equations and can automatically extract features from data to achieve precise model regression. In light of the high level of the uncertainty and limited data volume in MMG systems, reinforcement learning (RL) is suitable for real-time decision-making under complex and variable operating conditions [19]. In this way, Ref. [20] utilizes a fast online algorithm to solve the household load dispatching model, and the research results have achieved satisfactory results. Ref. [21] established a MG dispatch model considering renewable energy and used a hierarchical online algorithm to optimize the constructed objective function. However, the online optimization algorithm used in the above work has poor generalization performance compared to RL. Furthermore, the scheduling decision of the RL algorithm takes into account the potential impact of future long-term benefits, overcoming short-sightedness. However, existing RL methods are typically based on single-agent decision-making, which has limitations when dealing with complex scenarios. This is because single-agent RL relies on centralized scheduling and lacks autonomous learning capabilities, and may also face difficulties with the curse of dimensionality in complex multi-entities decision-making, resulting in convergence issues.

To address the aforementioned challenges, this paper proposes the use of multi-agent deep reinforcement learning (MADRL) for MMG optimal scheduling. Existing work has adopted MADRL to solve problems in power systems [22, 23, 24]. Ref. [25] proposes a layered hybrid MADRL to optimize a multi-service delivery business model that involves the coordination of multiple electric vehicles. Ref. [26] proposes a MADRL method for finding the optimal energy-saving strategy for hybrid electric vehicles. Ref. [27] utilizes the hybrid action space of MADRL to optimize off-grid building energy systems. Ref. [28] proposes the use of MADRL for the optimal scheduling of electric vehicle charging. However, the above works do not exploit the potential of MADRL to consider transactions between different entities. Besides, the generalization performance of the algorithm is also critical for practical applications. Ref. [29] uses the Nash-Q algorithm for multi-channel network system security control. However, this algorithm still faces the curse of dimensionality when dealing with complex scheduling scenarios, and its generalization ability is relatively poor, which limits its applicability. Moreover, this algorithm employs a discrete action space, which can lead to a reduction in calculation accuracy. In this regard, a soft actor-critic (SAC) algorithm, which combines value and policy iteration, has been successfully applied to power systems [30, 31]. To tackle the energy management problem of a MMG system, which consists of multiple renewable energy microgrids belonging to different operating entities, we propose a multi-agent soft actor-critic (MASAC) algorithm that leverages automated machine learning (AutoML) as a skill to improve the generalization of MASAC and utilizes experience replay buffer to reduce the temporal correlation between samples and improve training stability.

In summary, this paper proposes a MMG collaborative optimization scheduling model for MMG based on AutoML and MADRL to address the complex characteristics of the MMG system.

The main contributions of this study are summarized as follows:

- To address the issue of the transaction and complementarity of electric energy among multi-microgrids, we constructed a collaborative optimization scheduling model for MMG based on a multi-agent centralized training distributed execution framework. This model effectively facilitates energy transactions between different entities and reduces the MMG system operating cost.
- To enhance the generalization performance of the algorithm to cope with renewable energy uncertainties, we proposed an AutoML-based MASAC analysis method for MMG energy management. This approach eliminates the reliance on mathematical probability distributions for renewable energy outputs and increases the adaptability of the method to complex MMG scenarios.
- Simulation tests have demonstrated that the proposed method can effectively manage the demand between different microgrids and promote the consumption of renewable energy, while achieving power complementarity. Moreover, the proposed method has better economy and computational efficiency than other RL algorithms.

The remaining sections of this paper are organized as follows: Section 2 mainly introduces the MMG energy management model; while Section 3 presents the solution method of the proposed model. In Section 4, we conduct a comprehensive case analysis to demonstrate the effectiveness of the proposed method. Finally, Section 5 summarizes the paper.

## 2. Multi-Microgrid Energy Management Model

The MMG system studied in this paper comprises multiple microgrids connected to the distribution network. Each individual MG can interact with other microgrids and trade energy with the distribution network through transmission lines. Before introducing the scheduling model of multi-microgrid in this section, we first discuss the individual MG model in detail.

### 2.1. Optimal Modeling of the Individual Microgrid

To clearly demonstrate the MG model, Figure 1 shows a schematic diagram of an individual MG's structure, which is mainly composed of wind turbine (WT) units, photovoltaic (PV) units, electricity storage devices (ESD), micro gas turbines (MGTS), load unit and an energy management center. Furthermore, the energy management center is responsible for the energy management of the MG.

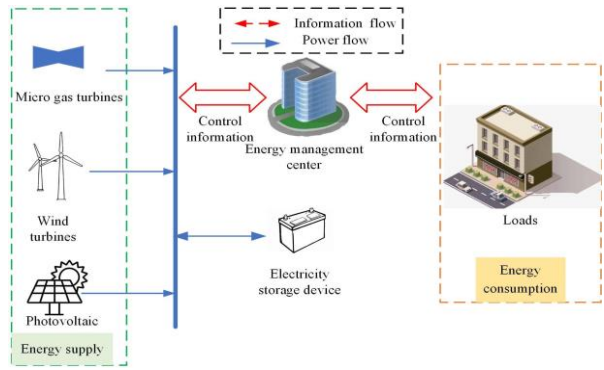

**Figure 1.** Schematic diagram of individual MG structure.

2.1.1. Distributed Generation

The distributed generation in the studied MG includes WT and PV units. Following the principle of data-driven dispatch, this study employs real wind and photovoltaic power generation data for subsequent analysis, instead of modeling wind power and photovoltaic power generation output using explicit expressions [32].

2.1.2. Micro Gas Turbines

The MGTS mainly burns natural gas to generate electricity, which can offer advantages such as high controllability and good power supply reliability. For the convenience of analysis, the cost function of the MGTS of MG $i$ is set as follows:

$$C_i^t(P_{MGTS,t,i}) = \lambda_{MGTS,MG\ i} P_{MGTS,t,i} \quad \forall t \tag{1}$$

$$P_{MGTS,i}^{\min} \leq P_{MGTS,t,i} \leq P_{MGTS,i}^{\max} \quad \forall t \tag{2}$$

where $C_i^t(P_{MGTS,t,i})$ represents the operating cost of the MGTS of MG $i$ at time $t$; $P_{MGTS,t,i}$ is the power generation of the MGTS of MG $i$ at time $t$; $\lambda_{MGTS,MG\ i}$ represent the power generation cost coefficients of MGTS of MG $i$; $P_{MGTS,i}^{\min}$ and $P_{MGTS,i}^{\max}$ are the minimum and maximum power generation of the MGTS of MG $i$, respectively.

2.1.3. Electricity Storage Devices

The ESD mainly achieves reasonable energy distribution by storing and releasing through the storage and release of electric energy, ultimately reducing the operating cost of the system [33]. At time $t+1$, the relationship between the available capacity of the ESD and its charge and discharge power of MG $i$ is expressed as:

$$S_{ESD,t+1,i} = S_{ESD,t,i} + \left(\eta_{ch} P_{ch,t,i} - P_{dc,t,i}/\eta_{dc}\right)\Delta t \quad \forall t \tag{3}$$

where $\eta_{ch}$ and $\eta_{dc}$ represent the charging and discharging rate of the ESD, respectively; $P_{ch,t,i}$ and $P_{dc,t,i}$ are the charging and discharging power of the ESD of MG $i$ in period $t$; $S_{ESD,t,i}$ and $S_{ESD,t+1,i}$ represent the capacity value of the ESD of MG $i$ in period $t$ and $t+1$. Furthermore, we define the state of charge $SOC_{ESD,t,i}$ of the ESD of MG $i$ at time $t$ to detect the capacity of the ESD in real time as follows:

$$SOC_{ESD,t,i} = S_{ESD,t,i} / S_{ESD,\max} \quad \forall t \qquad (4)$$

where $S_{ESD,\max}$ is the maximum capacity of the ESD. Besides, ESD improves the system economy through reasonable charge and discharge. For the convenience of analysis, the ESD operation and maintenance cost is set as follows:

$$C_{ESD,i}(t) = (|P_{ch,t,i}| + |P_{dc,t,i}|)\lambda_b \quad \forall t \qquad (5)$$

where $C_{ESD,i}(t)$ is the ESD operation and maintenance cost of MG $i$ at time $t$; $\lambda_b$ is the operation and maintenance cost coefficient per unit power.

*2.2. Multi-Microgrid Energy Management Model*

To provide a clear representation of the MMG model, Figure 2 illustrates a schematic diagram of the MMG structure, which comprises multiple microgrids connected to the distribution network. The electricity trading process between MGs as well as between MGs and the distribution network is based on energy price information to ensure the economical operation of the MMG. Besides, the unified control center is responsible for managing and integrating MMG price information and system power requirements, which are then sent to individual MG. And the energy management center is responsible for the energy management of the MG based on the information provided by the unified control center.

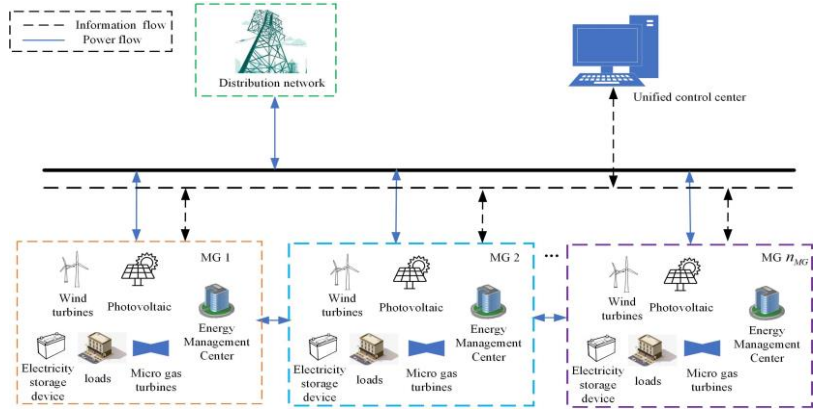

**Figure 2.** Schematic diagram of MMG structure.

2.2.1. Objective Function

The primary objective of MMG is to minimize the system's operating cost. The operating cost of MMG consists primarily of the operating cost of micro gas turbines, the transaction cost between MGs, the transaction cost between MGs and the distribution network, the operation and maintenance cost of ESD, the cost of active power loss, and the penalty cost of power imbalance between energy supply and consumption. Therefore, the objective function of MMG is:

$$\min \text{Cost} = \sum_{t=1}^{T} \sum_{i=1}^{n_{MG}} \left( C_i^t(P_{MGTS,t,i}) + C_{MG,i}(t) + C_{Grid,i}(t) + C_{ESD,i}(t) + \lambda_{loss} P_{loss,t,i} + \ell(P_{gap,t,i})^2 \right) \qquad (6)$$

where $n_{MG}$ is the number of MGs; $T$ is the number of time periods in a day; $C_i^t(P_{MGTS,t,i})$ is the operating cost of MGTS of MG $i$ at time $t$; $C_{MG,i}(t)$ is the transaction cost between the MG $i$ and other MGs at time $t$; $C_{Grid,i}(t)$ is the cost of transactions between MG $i$ and distribution network at time $t$; $C_{ESD,i}(t)$ is the ESD operation and maintenance cost of MG $i$ at time $t$; $\lambda_{loss} P_{loss,t,i}$ is the loss cost of MG $i$ in the process of energy transmission and generation side unit generation at time $t$; $\lambda_{loss}$ is the unit loss cost coefficient; $P_{loss,t,i}$ is the total power loss value of MG $i$ during the energy transmission process and generation side unit generation at time $t$; $\ell(P_{gap,t,i})^2$ is the penalty cost of MG $i$ in case of imbalance between energy supply and consumption at time $t$; $P_{gap,t,i}$ is the power difference of MG $i$ between the energy supplied and the energy consumed at time $t$; $\ell$ characterizes the penalty factor for the imbalance between energy supply and consumption.

2.2.1.1. Transaction Cost Between Microgrids

The transaction cost between MGs is mainly determined by the price of electricity transacting between MGs as well as the amount of electricity traded. To reasonably arrange the transaction energy between MGs, the transaction cost of MG $i$ is expressed as:

$$C_{MG,i}(t) = \sum_{j=1,j\neq i}^{n_{MG}} \delta_{MG,t} P_{ij,t} \quad \forall t \tag{7}$$

where $\delta_{MG,t}$ is the purchase and sale price of electricity between MGs during the period $t$, and it is stipulated that the purchase price is equal to the electricity sale price; $P_{ij,t}$ is the transaction power between MGs $i$ and $j$ during period $t$, when the value is greater than 0, it is electricity purchase, and when it is less than 0 for electricity sales.

2.2.1.2. Transaction Cost of MGs and Distribution Network

To reasonably arrange the electricity traded between the MGs and the distribution network as well as reduce the pressure on the power supply of the grid, the following transaction cost of MG $i$ are set:

$$C_{Grid,i}(t) = \delta_{Grid,t} P_{ig,t} \quad \forall t \tag{8}$$

where $\delta_{Grid,t}$ is the purchase and sale price of electricity between the MGs and distribution network during period $t$, and the stipulated electricity purchase price is greater than the electricity sale price, besides, the purchase and sale price between the MGs is between the purchase price and the sale price of the MGs and the distribution network; $P_{ig,t}$ is the electricity traded between the MG $i$ and the distribution network during period $t$, if the value is greater than 0, it represents electricity purchase; on the contrary, it represents electricity sales.

To ensure the interests of the distribution network and to encourage energy transactions among MGs, we set the purchase and sale price of transactions between MGs to be lower than the purchase price between MGs and the distribution network, besides, the purchase and sale price between the MGs is between the purchase price and the sale price of the MGs and the distribution network. In case the MG experiences a shortage of power, it gives priority to purchasing power from other MGs. If the demand is still not met, the MG purchases power from the grid. Similarly, when the MG has surplus power and other MGs face a power shortage, it prioritizes meeting the load demand in the MMG system.

2.2.1.3. Microgrid Power Loss

The power loss taken into account in this study refers to the active power loss that occurs during the power generation of the generator set and energy transmission process. The generator-side unit comprises MGTS, WT, and PV. The formula used to calculate the specific power loss is as follows:

$$P_{loss,t,i} = \psi_{MGTS,t,i} P_{MGTS,t,i} + \psi_{PV,t,i} P_{PV,t,i} + \psi_{WT,t,i} P_{WT,t,i} \quad \forall t \tag{9}$$

$$\psi_{MGTS,t,i} = \frac{\partial P_{loss,t,i}}{\partial P_{MGTS,t,i}} \quad \forall t \tag{10}$$

$$\psi_{PV,t,i} = \frac{\partial P_{loss,t,i}}{\partial P_{PV,t,i}} \quad \forall t \tag{11}$$

$$\psi_{WT,t,i} = \frac{\partial P_{loss,t,i}}{\partial P_{WT,t,i}} \quad \forall t \tag{12}$$

where $\psi_{MGTS,t,i}$, $\psi_{PV,t,i}$, $\psi_{WT,t,i}$ represent the power loss coefficients of micro gas turbines, photovoltaics, and wind turbines respectively; $P_{WT,t,i}$ is the power generated by the WT of MG $i$ at time $t$; $P_{PV,t,i}$ is the power generated by PV of MG $i$ at time $t$.

2.2.1.4. Power Imbalance Between Energy Supply and Consumption

To facilitate the integration of renewable energy sources and achieve a balance between energy supply and demand, the unbalanced power of MG $i$ is set to:

$$P_{gap,t,i} = P_{sup,t,i} - P_{con,t,i} \quad \forall t \tag{9}$$

$$P_{sup,t,i} = P_{MGTS,t,i} + P_{WT,t,i} + P_{PV,t,i} + P_{dc,t,i} + P_{ij,t} + P_{ig,t} \quad \forall t \tag{10}$$

$$P_{con,t,i} = P_{load,t,i} + P_{ch,t,i} + P_{loss,t,i} \quad \forall t \tag{11}$$

where $P_{sup,t,i}$ is the energy provided by MG $i$ at time $t$; $P_{con,t,i}$ is the energy consumed by MG $i$ at time $t$; $P_{load,t,i}$ is the load power of MG $i$ at time $t$.

2.2.2. Constraints

2.2.2.1. Electrical Balance Constraint

To reasonably adjust the output of the power generation side and maintain the balance of energy supply and demand in the system, we set the following energy balance constraint of MG $i$:

$$P_{MGTS,t,i} + P_{WT,t,i} + P_{PV,t,i} + P_{dc,t,i} + P_{ij,t} + P_{ig,t} = P_{load,t,i} + P_{ch,t,i} + P_{loss,t,i} \quad \forall t \tag{12}$$

2.2.2.2. Constraints of Electricity Storage Devices

To ensure that the charging and discharging power of ESD is within the allowable range, the limiting conditions of MG $i$ are as follows [34]:

$$0 \leq P_{ch,t,i} \leq P_{ch,\max} \quad \forall t \tag{13}$$

$$0 \leq P_{dc,t,i} \leq P_{dc,\max} \quad \forall t \tag{14}$$

where $P_{ch,\max}$ and $P_{dc,\max}$ represent the maximum charging and discharging power of ESD.

To ensure that the ESD capacity is within the allowable range, the capacity of MG $i$ must meet the following limits:

$$S_{ESD,\min} \leq S_{ESD,t,i} \leq S_{ESD,\max} \quad \forall t \tag{15}$$

where $S_{ESD,\min}$ is the minimum capacity of ESD.

Start and end limits: To ensure that the initial conditions remain consistent for each scheduling cycle, the ESD should adhere to the following start and end limits:

$$S_0 = S_{T,end} = S_{ESD,\min} \tag{20}$$

where $S_0 = 0$ and $S_{T,end}$ are the capacity of ESD at the beginning and end of the scheduling period $T$ (in this work, $T$ is taken as 24 hours).

2.2.2.3. Constraints on Power Trading Between MGs and Distribution Network

To avoid the excessive purchase of electricity from the distribution network, which may lead to higher electricity costs, the electricity traded between the MG $i$ and the distribution network at time $t$ is set as:

$$-P_{ig,\max} \leq P_{ig,t} \leq 0 \quad P_{ig,t} \leq 0 \quad \forall t \tag{21}$$

$$0 \leq P_{ig,t} \leq P_{ig,\max} \quad P_{ig,t} \geq 0 \quad \forall t \tag{22}$$

where $P_{ig,\max}$ is the maximum power when the MG $i$ trades with the grid.

2.2.2.4. Constraints on Electricity Traded Between Microgrids

To prevent excessive power trading between MGs as well as avoid causing a supply-demand imbalance in MGs, we set the following power trading constraints:

$$P_{ij,t} = -P_{ji,t} \quad \forall t \tag{23}$$

$$0 \leq P_{ij,t} \leq P_{ij,\max} \quad \forall t \tag{16}$$

where $P_{ji,t}$ is the electricity traded between MG $j$ and MG $i$ at time $t$; $P_{ij,\max}$ is the maximum power traded between MG $i$ and MG $j$.

## 3. Model Solving

In this section, we first describe in detail the automated machine learning used to improve the generalization of MASAC, followed by a detailed introduction to the MASAC methodology proposed in this study.

### 3.1. Automated Machine Learning

Typically, the process of selecting neural network structures and hyperparameters for machine learning models involves a trial-and-error approach, which can be both tedious and challenging. To overcome this issue, we propose the use of complex control structures to operate machine learning models that can automatically learn appropriate parameters and configurations without the need for human intervention [35, 36, 37].

Optimizing hyperparameters for DRL algorithms is widely acknowledged as a complex task. In this study, we tackle this challenge by utilizing the currently popular AutoML technique to automatically find the best combination of hyperparameters for DRL. Figure 3 illustrates the structure of our approach. We use the metis tuner algorithm [38, 39] to optimize hyperparameters. By leveraging metis to predict the next trial instead of guessing randomly, the AutoML finds the best hyperparameters for DRL. Moreover, we utilize AutoML to optimize the hyperparameters of MASAC with discount factor $\gamma$, actor network learning rate $a\_l$, critic network learning rate $c\_l$, mini-batch $N$, and adjustment coefficient $\partial$.

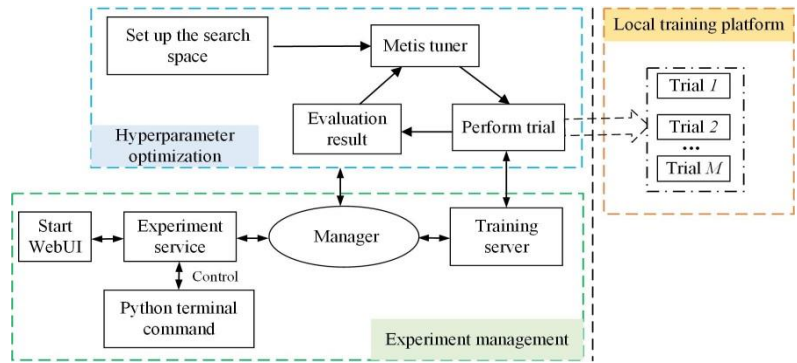

**Figure 3.** AutoML-based MASAC hyperparameter optimization.

Moreover, metis uses latin hypercube sampling (LHS) in stratified sampling [40], which divides the range of $U$ parameters into $D$ intervals and picks data points from the interval one at a time. Therefore, the number of combinations $C$ of bootstrapping trials is

$$C = \left( \prod_{d=0}^{D-1} D - d \right)^{U-1} \tag{17}$$

After obtaining the above number of combinations, metis iteratively trains the Gaussian process model to enhance the robustness of tuning.

### 3.2. MASAC Methodology

Generally, the MADRL task can be described as a markov decision process game (MDP game) [41]. Specifically, the MDP game consists of five key elements $\{[S_i]_{n_{MG}}, [A_i]_{n_{MG}}, [\rho_i]_{n_{MG}}, [R_i]_{n_{MG}}, [\gamma_i]_{n_{MG}}\}$, where $S_i$ is the state set of agent $i$, $[S_i]_{n_{MG}}$ is the state set of all agents; $A_i$ represents the action set of agent $i$, and $[A_i]_{n_{MG}}$ is the action set of all agents; $\rho_i$ is the state transition matrix of agent $i$, $[\rho_i]_{n_{MG}}$ is the set of state transition matrices of all agents; $R_i$ is the return reward of agent $i$ from state $S_{i,t-1}$ to state $S_{i,t}$, $[R_i]_{n_{MG}}$ is the reward set of all agents; $\gamma_i$ is the discount factor of agent $i$, which will affect the convergence of the algorithm, $[\gamma_i]_{n_{MG}}$ is the discount factor set of all agents. During the training process, the agent optimizes its own strategy, and the accumulated reward value gradually increases and tends to stabilize.

In this section, we apply the MDP game to the MMG scheduling model in this research. The key elements for each MG $i$ are as follows.

1) Agent: in each MG, the energy management center is set as an agent for the DRL algorithm.

2) Environment: the environment is composed of PV, WT, ESD, loads, distribution network, and micro gas turbines.

3) State: the state is used to describe the environmental feedback of the action taken by the agent in the current environment. Specifically, the state includes the load power $P_{load,t,i}$ of the MG $i$ in period $t$, the state of charge $SOC_{ESD,t,i}$ of the electricity storage device of the MG $i$ in period $t$, the power generation $P_{WT,t,i}$ of the WT of MG $i$ in period $t$, the power generation $P_{PV,t,i}$ of the PV of MG $i$ in period $t$, the transaction price $\delta_{MG,t}$ between MGs in period $t$, as well as the transaction price information $\delta_{Grid,t}$ between MGs and distribution network in period $t$. Therefore, the state set of agent $i$ is:

$$S_i = \{P_{load,t,i}, SOC_{ESD,t,i}, P_{WT,t,i}, P_{PV,t,i}, \delta_{MG,t}, \delta_{Grid,t}\} \tag{18}$$

4) Action: action is mainly composed of the output $P_{MGTS,t,i}$ of the MGTS of the MG $i$ at time $t$, the transaction strategy $P_{ij,t}$ between MG $i$ and MG $j$ at time $t$, and the transaction strategy $P_{ig,t}$ between MG $i$ and the distribution network at time $t$. Therefore, the action set of agent $i$ is:

$$A_i = \{P_{MGTS,t,i}, P_{ij,t}, P_{ig,t}\} \tag{19}$$

5) Reward: the cost of each MG $i$ includes the operation cost of micro gas turbines, transaction cost with other MGs, transaction cost with the distribution network, ESD operation and maintenance cost, active power loss cost, and unbalanced power penalty cost. Besides, the goal of MG is to minimize the operating cost, therefore, the reward of agent $i$ at time $t$ is defined as follows:

$$R_i(t) = -(C_i^t(P_{MGTS,t,i}) + C_{MG,i}(t) + C_{Grid,i}(t) + C_{ESD,i}(t) + \lambda_{loss} P_{loss,t,i} + \ell(P_{gap,t,i})^2) \quad \forall t \tag{20}$$

In this regard, we define the total reward as the sum of the reward values of all agents. Moreover, in complex multi-agent interactive scenarios, the policy gradient method of the general single-agent reinforcement learning algorithm tends to increase variance with the number of agents. Besides, most single-agent reinforcement learning is a centralized learning method, which is not scalable. However, multi-agent reinforcement learning algorithms demonstrate superiority in multi-agent interaction scenarios. They acquire additional information during training to enhance stability, and the specific execution of the strategy depends only on the observation of the agent itself, without relying on the additional information. In this study, we adopt the MASAC algorithm, which follows the basic idea of centralized training and distributed execution (CTDE). Specifically, during training, the algorithm incorporates a global critic to guide actor training, while during testing, only actors with local observation environments are used to take action [42]. The advantage of this method is that it improves the efficiency of learning during training and improves the stability of training in a multi-agent environment. In this way, the framework of CTDE is shown in Figure 4.

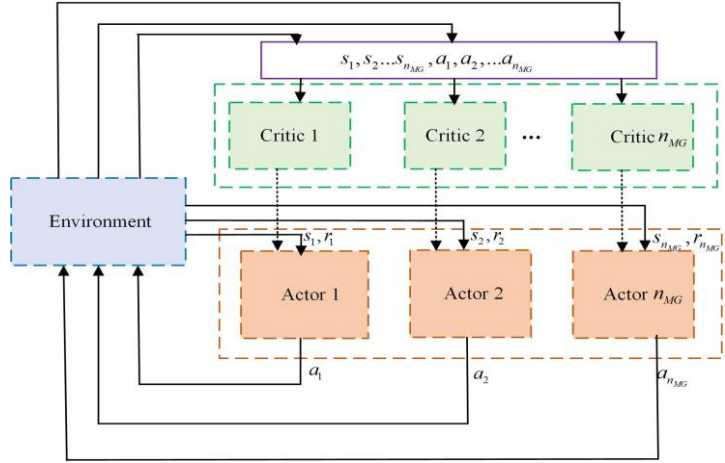

**Figure 4.** CTDE structure diagram.

Furthermore, we utilize the CTDE framework to extend the SAC algorithm to the dispatching scenario of multi-agent microgrids, which we call MASAC. This approach allows for the training of multiple agents in a high-dimensional continuous action space. The goal of MASAC is to maximize exploration by increasing the entropy, thereby avoiding falling into a local optimal solution, and finding the strategy of global maximization. In MASAC, the actor of agent $i$ updates the parameters of the policy network according to the gradient descent theory. The objective function of the specific policy network is as follows:

$$J(\varphi_i)_{\pi_i} = E_{x \sim \Re}[\kappa \log(\pi_{\varphi_i}(a_i | s_i)) - Q_{\xi,i}(x,a)] \tag{21}$$

$$a = \{a_1, a_2, \ldots a_{n_{MG}}\} \tag{30}$$

$$x = \{s_1, s_2, \ldots s_{n_{MG}}\} \tag{31}$$

$$a = \{a_1, a_2, \ldots a_{n_{MG}}\} \tag{32}$$

$$r = \{r_1, r_2, \ldots r_{n_{MG}}\} \tag{33}$$

$$x' = \{s'_1, s'_2, \ldots s'_{n_{MG}}\} \tag{34}$$

where $\pi_{\varphi_i}$ is the parameter of the actor network $\pi$ of each agent $i$ is $\varphi$; $Q_{\xi,i}$ is the parameter of the critic network of agent $i$ is $\xi$; $\kappa$ is the temperature parameter, which is used to control the influence ratio of entropy and reward; $\Re$ is the experience replay buffer, which is mainly used to store the joint state $x$, action $a$, reward $r$, and next state $x'$; $a$ is the action input into the critic network. Besides, the critic network of agent $i$ updates the parameter $\xi$ by minimizing the bellman error $J(\xi_i)_Q$:

$$J(\xi_i)_Q = E_{x,a,r,x' \sim \Re}[\frac{1}{2}(Q_{\xi,i}(x,a) - w)^2] \tag{22}$$

$$w = r_i + \gamma E[Q_{\overline{\xi}}(x',a') - \kappa \log(\pi_{\varphi_i}(a_i' | s_i'))] \tag{23}$$

where $\overline{\xi_i}$ is the target critic network parameter of agent $i$; $a'$ is the next action of agent $i$. During training, the actor and current critic network are utilized, while the target critic network performs parameter transfer from the current network to stabilize the training effect. After updating each critic network parameter, the target critic network parameter is soft updated, as follows:

$$\overline{\xi_i} = \phi \xi_i + (1-\phi)\overline{\xi_i} \tag{24}$$

where $\phi$ is the hyperparameter controlling the soft update. Moreover, one of the main features of MASAC is the regularization of policy entropy. By increasing the exploration of actions, it can speed up the speed of the train and improve the quality of learning, preventing the policy from prematurely converging to a bad local optimal solution.

Moreover, the effective utilization of sampled data is also a key issue in MASAC. Experience replay buffer is a commonly used technique to store old and new experiences to prevent temporal correlations among samples [43], thereby improving the efficiency and quality of learning during training.

Based on the above analysis, Algorithm 1 summarizes the final MASAC algorithm.

---

**Algorithm 1: MASAC Algorithm Based on AutoML for Multi-Microgrid Optimal Scheduling**

1: Initialize the neural network parameters $\varphi$ and $\xi$ of actor and critic.
2: Initialize the replay buffer $\Re$ with size $S_{\Re}$.
3: **for** trial = 1 : $M$ **do**
4:    Select a set of hyperparameters from the search space according to the Metis Tuner.
5:    **for** episode = 1 : $E$ **do**
6:      Select random action from the action space.
7:      Select the initial state from the state space.
8:      **for** $t$ = 1 : $H$ **do**
9:         Each agent $i$ selects action $a_i$ from the action space.
10:        Interact joint actions $a = \{a_1, a_2, ... a_{n_{MG}}\}$ with the environment to get corresponding states $x'$ and rewards $r$.
11:        Store transition $(x, a, r, x')$ in experience replay buffer $\Re$.
12:        **for** agent = 1: $n_{MG}$ **do**
13:           Sample a mini-batch of $N$ experience $\left(x^N, a^N, r^N, x'^N\right)$ from the experience replay buffer $\Re$.
14:           Updating the critic network by minimizing the loss function.
15:           Updating the actor network via gradient descent.
16:        **end for**
17:        Updating critic target network parameters using soft update.
18:      **end for**
19:    **end for**
20:    Collect the reward and upload it to the Metis Tuner.
21: **end for**
22: Select the best hyperparameters and policies.

---

*3.3. Solving process*

The solution process of the MMG scheduling model is as follows:
Step 1: Construct a scheduling model according to the formula (6)-(24).
Step 2: Input the MMG parameters.
Step 3: Set and update the episode of MASAC training.
Step 4: According to the state set and action set, calculate the reward function according to the formula (28).
Step 5: Determine whether a solution exists. If it exists and meets the stopping criteria, the process terminates; otherwise, return to Step 2.
Step 6: Obtain the optimal scheduling strategy for MMG.

**4. Case Study**

To verify the effectiveness of the proposed scheduling model and method, the following simulation experiments are carried out. Moreover, the MASAC algorithm proposed in this study has been implemented in Python 3.8 using Pytorch 1.10. All simulation tests are carried out on a PC platform equipped with Intel Core i5-6300HQ CPU (2.3 GHz) and 8GB RAM.

*4.1. Settings in Test Case*

In this study, we set up a test case of a MMG system test case consisting of two microgrids. The key components of the MMG system include micro gas turbines, wind turbines, photovoltaics, electricity storage devices, loads, and energy management centers. Besides, the data records for WT power generation are provided by Fortum Oyj from a wind farm in Finland, while the data related to PVs are obtained from [44]. Moreover, the time range of the simulation test is set to *T*=24 hours, and the time interval is *t*=1 hours. Figure 5 and Figure 6 are the data curves of WT and PV power generation and load power of MG 1 and MG 2, respectively. It can be seen from the data that the load value of MG 1 is higher than the output of renewable energy in multiple periods, while the load value of MG 2 is lower than the output of renewable energy in multiple periods. Figure 7 depicts the price information for transactions between the MGs and the grid as well as between MGs. Moreover, the maximum trading power between the MGs and the grid is 500 kW, and the maximum trading power between MGs is 200 kW. Table 1 describes the main parameter settings of the MG [45]. Besides, the main implementation details of MASAC are as follows. Specifically, the basic structure of the neural network of each MG is consistent, and the Adam optimizer is used. The $\gamma$ of MASAC is 0.916, the learning rates of actor and critic are $a\_l$=0.0004 and $c\_l$=0.0006 respectively, the size of the experience replay buffer $S_{\Re}$ is 10000, $\partial$ is 0.159, and the sampling mini-batch *N* is 512.

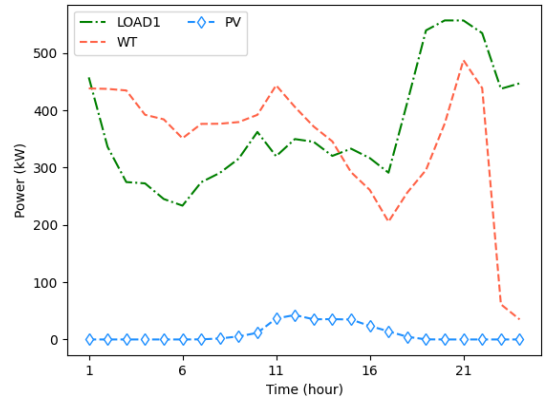

**Figure 5.** WT, PV output and load power curve of MG 1.

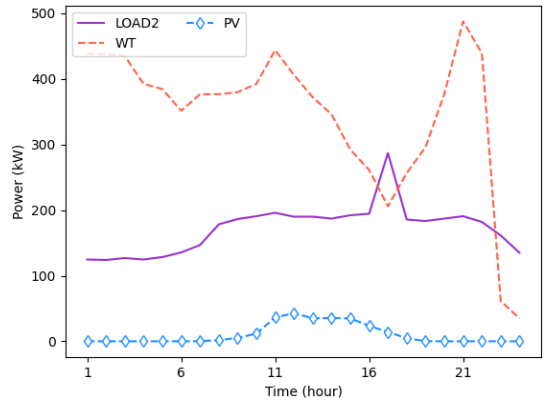

**Figure 6.** WT, PV output and load power curve of MG 2.

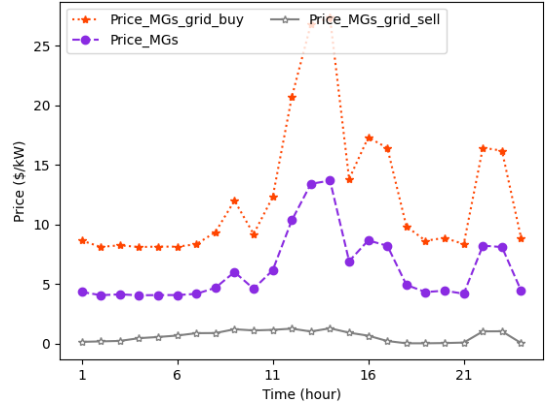

**Figure 7.** Transaction prices between MGs and grid as well as between MGs.

**Table 1.** Main parameter settings of each MG.

| Parameter | Value | Parameter | Value |
| --- | --- | --- | --- |
| $P_{ch,\max}$ (kW) | 100 | $\lambda_{MGTS,MG\ 1}$ ($/kWh) | 1.3 |
| $P_{dc,\max}$ (kW) | 100 | $\lambda_{MGTS,MG\ 2}$ ($/kWh) | 1.5 |
| $S_{ESD,\max}$ (kWh) | 200 | $\lambda_{loss}$ ($/kWh) | 1.35 |
| $\eta_{ch}, \eta_{dc}$ | 0.9 | $\ell$ | 0.5 |
| $P_{MGTS,i}^{\min}$ (kW) | 5 | $n_{MG}$ | 2 |
| $P_{MGTS,i}^{\max}$ (kW) | 30 | $\psi_{MGTS,t,i}, \psi_{PV,t,i}, \psi_{WT,t,i}$ | 0.02 |
| $\lambda_b$ ($/kWh) | 0.5 | | |

*4.2. Results and Analysis*

4.2.1. Analyze Optimization Results Using AutoML

To evaluate the effectiveness of AutoML, the following simulation tests have been carried out. AutoML assesses the intermediate results generated by the current hyperparameter selection and offers reasonable suggestions for the next hyperparameter trajectory. Finally, the hyperparameter selection results for each trajectory are displayed on the WebUI, as shown in Figure 8.

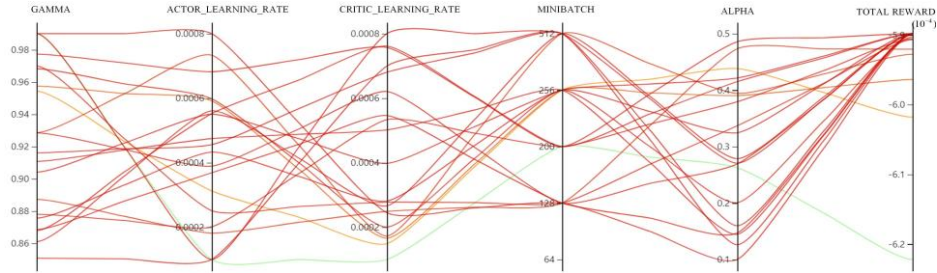

**Figure 8.** Results of hyperparameter optimization using AutoML.

Figure 8 shows the results of AutoML's optimization of the hyperparameters required by MASAC. In this figure, each curve represents a set of hyperparameters for a trial, each ordinate represents a range of hyperparameters, and the last ordinate is the reported total reward value for all agents using those hyperparameters. Besides, the darker the red, the more appropriate the hyperparameter set, and the green color indicates that the selected combination of hyperparameters cannot achieve satisfactory results.

Besides, to validate the rationality of AutoML's multiple trial results, the following simulations were conducted. Figure 9 shows the results of the final report total reward value for all agents in multiple trials using AutoML, where each coordinate point represents a trial. It can be seen from the figure that except for a few trial results that deviate from the normal value of the trial, most of the trials can achieve satisfactory results. Based on the final reported total reward value for all agents, it is evident that the designed AutoML can achieve desirable optimization results.

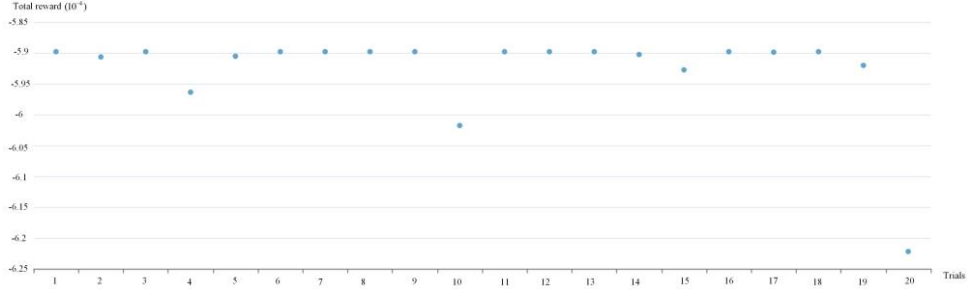

**Figure 9.** Total reward distribution under multiple trials.

Upon analyzing the results above, it is evident that AutoML is capable of selecting the optimal combination of hyperparameters for MASAC, leading to an improvement in the algorithm's generalization ability and learning efficiency.

Besides, to verify the adaptability of AutoML to parameter adjustment when the input parameters are changed, we changed the values of the input parameters in Table 2 and performed simulation experiments. Note that Table 2 here is only a set of parameters for verifying the adaptive setting of the above-mentioned AutoML, which is different from the parameters of the MMG optimization model constructed in Table 1. Besides, we set up two experiments as follows:

**Table 2.** A set of parameters to verify the adaptability of AutoML.

| Parameter | Value | Parameter | Value |
|---|---|---|---|
| $P_{ch,\max}$ (kW) | 100 | $\lambda_{MGTS,MG\ 1}$ ($/kWh) | 0.1 |
| $P_{dc,\max}$ (kW) | 100 | $\lambda_{MGTS,MG\ 2}$ ($/kWh) | 0.2 |
| $S_{ESD,\max}$ (kWh) | 200 | $\lambda_{loss}$ ($/kWh) | 0.15 |
| $\eta_{ch}, \eta_{dc}$ | 0.9 | $\ell$ | 0.5 |
| $P_{MGTS,i}^{\min}$ (kW) | 5 | $n_{MG}$ | 2 |
| $P_{MGTS,i}^{\max}$ (kW) | 30 | $\psi_{MGTS,t,i}, \psi_{PV,t,i}, \psi_{WT,t,i}$ | 0.02 |
| $\lambda_b$ ($/kWh) | 0.06 | | |

Experiment 1: Optimizing the scheduling model under the input parameters in Table 1.
Experiment 2: Optimizing the scheduling model under the input parameters in Table 2.

Table 3 shows the best hyperparameter results optimized by AutoML for MASAC under different experiments. It can be seen from the table that under different input parameters, the optimal combination of hyperparameters optimized by the experiment is different. This shows that AutoML can automatically select the best hyperparameters for MASAC according to different inputs to formulate a reasonable scheduling strategy. And the AutoML adaptability was verified.

**Table 3.** Results of AutoML optimization hyperparameters under different experiments.

| | $\gamma$ | $a\_l$ | $c\_l$ | $N$ | $\partial$ |
|---|---|---|---|---|---|
| Experiment 1 | 0.916 | 0.0004 | 0.0006 | 512 | 0.159 |
| Experiment 2 | 0.877 | 0.0002 | 0.0007 | 128 | 0.269 |

4.2.2. Electrical Balance Analysis of Each MG

To verify the electrical balance effect of each MG, we conducted a simulation analysis, and the results are shown in Figure 10 and Figure 11. The analysis demonstrates that each MG attains its own energy supply balance by means of energy transactions with other MGs and the distribution network.

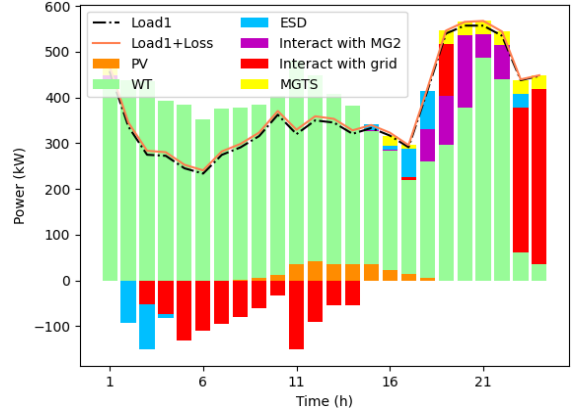

**Figure 10.** Electrical balance of MG 1.

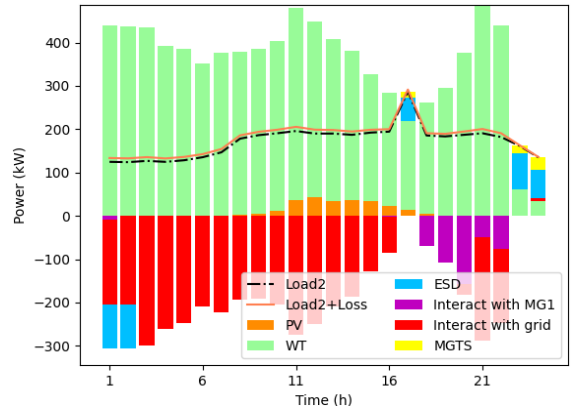

**Figure 11.** Electrical balance of MG 2.

It is evident from Figure 10. that the WT and PV outputs of MG 1 are lower than the power of the electric load during the peak period of power consumption, resulting in a power deficit for this MG. As a result, MG 1 receives additional power supply from other MGs and the distribution network to meet the demand. Similarly, Figure 11 shows that the WT and PV outputs of MG 2 are higher than the power of the electric load in most periods of time, indicating a power surplus MG. The excess electricity produced can be sold to other MGs or the distribution network, resulting in additional income. Specifically, during the off-peak period of electricity consumption, each MG charges the excess electricity to its own ESD, which is then discharged during the peak period of electricity consumption. Additionally, the excess electricity is sold to other MGs and the distribution network to generate further revenue. During peak hours of power consumption, MG 1 discharges its own ESD first. If the energy supplied by the MG itself is insufficient, it purchases electricity from other power surplus MGs at a lower transaction price than the price between MGs and the distribution network. If the energy demand is still not met, electricity is purchased from the grid to maintain the energy supply and demand balance. Through the energy complementarity between the MGs, the full utilization of energy is realized, and the power supply pressure of the distribution network is reduced. Additionally, the electrical balance of each MG is achieved.

4.2.3. Economic Analysis

In order to verify the effectiveness of the proposed multi-microgrid scheduling model, the following two modes are set and simulation experiments are carried out.

Model 1: Transactions between microgrids are not considered in a single MG, only transactions between microgrids and distribution network are considered.

Model 2: Consider transactions between microgrids as well as between microgrids and distribution network in a single MG.

Table 4 shows the operating cost of MMG in two different modes. Besides, the MMG in this study works in Mode 2. It can be seen from Table 4 that the cost of MMG under Mode 2 is reduced by 7.36% compared with that under Mode 1. This shows that the power interaction between microgrids can effectively reduce the operating cost of the MMG system and improve the economics of system operation.

Table 4. Operating cost of MMG under different modes.

|  | Model 1 ($) | Model 2 ($) |
|---|---|---|
| MMG | 63624.00 | 58942.00 |

4.2.4. Analysis of Transactions Between MGs as well as Between MGs and the Distribution Network

To verify the effectiveness of the proposed scheduling scheme, simulation results were tested and presented in Figure 12, which describes the process of energy transactions between MGs as well as between MGs and the grid. The simulation results reveal that the power purchase price between MGs is much lower than that of MGs and the grid, as well as the price of electricity sold between MGs is higher than the price of electricity sold between MGs and the grid, hence the priority of energy transactions between MGs is higher than that of MGs and the grid. Specifically, when MG 2 has sufficient power, it gives priority to selling the excess power to the power shortage MG 1, and then finally sells the remaining energy to the grid. Similarly, when the power shortage MG 2 cannot meet the energy demand, it gives priority to purchasing power from MG 1 and then from the grid. The above analysis shows that the MMG system optimally utilizes energy and achieves a high economic utilization of energy through the energy complementation of each MG.

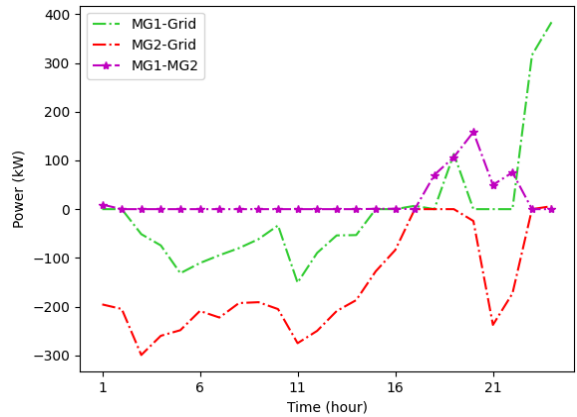

Figure 12. Electric power trading.

4.2.5. ESD Charging and Discharging Strategy Analysis

To verify the effectiveness of the ESD charging and discharging strategy, a simulation test was conducted and the results are shown in Figure 13. As can be observed from the figure, both MG 1 and MG 2's ESD store enough power when the power is sufficient, and when the power consumption peaks and the power is short, the ESD releases the stored energy. The power released by the ESD reduces the power purchased by the MG from the grid, further relieving the pressure on the grid power supply. And the charging and discharging strategy of ESD considers the energy shortage in future peak hours, effectively increasing MMG system operating flexibility.

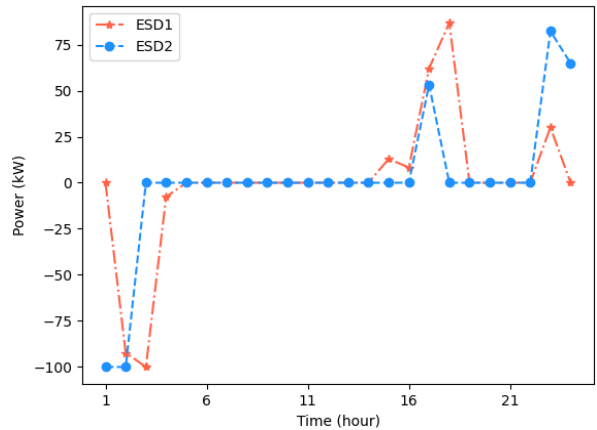

Figure 13. ESD charging and discharging strategy.

4.2.6. Performance Comparison with other RL Algorithms

To show the superiority of the proposed method, it has been tested in comparison with other RL methods. Figure 14 illustrates the convergence of different RL algorithms. During the initial learning phase, each RL method explores different directions randomly, which may not result in a more profitable policy and consequently leads to a lower total reward value for all agents. However, as the accumulated experience increases, the total reward value of all RL methods starts to increase continuously and eventually converges.

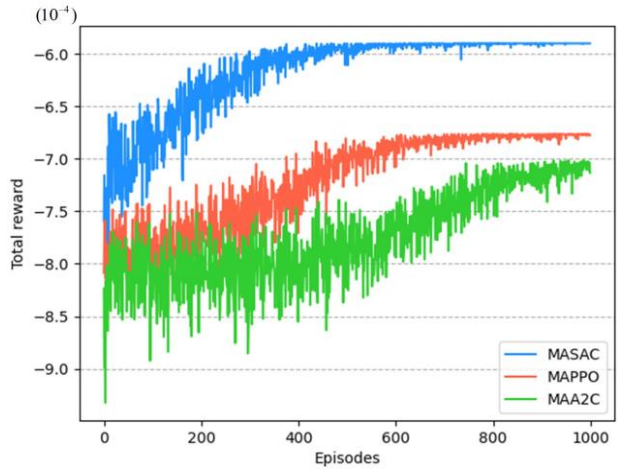

**Figure 14.** Comparison of different RL algorithm total reward.

Furthermore, the proposed method exhibits significantly higher total reward value for all agents compared to other RL algorithms, indicating a lower operating cost. Here, the proximal policy optimization (PPO) and advantage actor-critic (A2C) algorithms are extended to the multi-agent space in this study, named MAPPO and MAA2C respectively. Based on the results shown in Figure 14, it can be observed that the proposed method has reduced operating costs by 12.9% and 17.30% compared to MAPPO and MAA2C, respectively. The reasons for this are analyzed as follows: (1) The experience replay buffer and AutoML have improved the stability of training, and (2) MASAC has effectively enhanced the generalization performance through the framework of maximum entropy and CTDE. Hence, it can be concluded that our proposed method is more economical compared to other RL methods.

Additionally, to verify the convergence performance of the proposed method, we conducted simulation tests and present the results below. Table 5 compares the number of episodes and the calculation time required for different RL algorithms to reach convergence. Table 5 clearly indicates that the proposed method has the shortest convergence time and requires the least number of episodes to converge, as compared to other RL algorithms. Thanks to the strong generalization of the MASAC, it can quickly identify the optimal strategy and converge and stabilize faster. Thus, the proposed method outperforms other RL algorithms in terms of computational performance.

**Table 5.** Convergence comparison of different RL algorithms.

| Solution method | Number of episodes | Convergence time (s) |
| --- | --- | --- |
| Proposed method | 545 | 236.43 |
| MAPPO | 771 | 267.67 |
| MAA2C | 995 | 339.24 |

4.2.7. Analysis of Convergence and Computational Efficiency Across Multiple Runs

In order to verify the convergence performance and computational efficiency of the proposed method for multiple runs with the same input parameters, the following simulation experiments are performed.

Table 6 presents the total reward value for all agents and computation time in multiple runs with the same input parameters. Here, each run comprises 1000 episodes. It can be seen from the table that the total reward value for all agents has not changed across multiple runs, and the difference between the longest calculation time and the shortest calculation time during the training process is 12.01 seconds, which falls within an acceptable range. Additionally, the average computation time over 10 runs is 446.17 seconds. This shows that the proposed method can converge stably across multiple runs.

**Table 6.** Total reward value for all agents and computation time in multiple runs.

| Run number | Total reward ($10^{-4}$) | Computation time (s) |
|---|---|---|
| 1 | -5.89 | 442.00 |
| 2 | -5.89 | 449.11 |
| 3 | -5.89 | 451.78 |
| 4 | -5.89 | 447.33 |
| 5 | -5.89 | 444.08 |
| 6 | -5.89 | 439.77 |
| 7 | -5.89 | 447.56 |
| 8 | -5.89 | 439.99 |
| 9 | -5.89 | 449.42 |
| 10 | -5.89 | 450.68 |

## 5. Conclusion

To investigate the complementarity and trading of electric energy between MGs under different operating entities in MMG, this study proposes a multi-agent deep reinforcement learning scheduling method. The proposed method employs a multi-agent centralized training distributed execution framework to address uncertainty in the environment and determine the optimal trading strategy. Based on the simulation results, we draw the following conclusions.

(1) The established MMG scheduling model, based on a multi-agent centralized training distributed execution framework, allows the each MG to adjust the power interaction between MGs based on energy transaction prices and energy demand. This helps to reduce the cost of energy utilization and dependence on grid energy supply, while effectively facilitating energy transactions between different entities and improving the economics of MMG system operation.

(2) The developed MASAC algorithm, based on automated machine learning, has been shown to be capable of addressing the collaborative optimal scheduling of MMG and achieving satisfactory convergence by learning from historical experience. This approach is better suited to complex scheduling scenarios and real-time online scheduling decisions.

(3) The test results prove that the proposed method is more economical and computationally efficient than other RL algorithms.

In future work, various flexible loads will be considered to increase the flexibility on the demand side. It's interesting to extend this work to include the dispatch of multiple energy forms, such as heat and power, in an integrated energy system [46, 47]. Another topic worthy of research is the resilient scheduling of information attacks [48].